\documentclass[
  aps,
  prl,
  reprint,
  twocolumn,
  amsmath,amssymb,
  superscriptaddress,
  longbibliography
]{revtex4-2}

\usepackage{graphicx}
\usepackage{amsmath}
\usepackage{amssymb}
\usepackage{braket}
\usepackage{physics}
\usepackage[T1]{fontenc}
\usepackage[dvipsnames]{xcolor}
\usepackage[colorlinks,linkcolor=red,citecolor=blue,urlcolor=blue]{hyperref}

\newcommand{\kappap}{\kappa_+}
\newcommand{\kappam}{\kappa_-}
\newcommand{\epsv}{\varepsilon_v}

\begin{document}

\title{Electrically Tunable Valley-Based Qubits in Moiré Quantum Dots}

\author{Yasser Saleem}
\affiliation{
Institute for Theoretical Physics and Astrophysics,
and W\"urzburg-Dresden Cluster of Excellence on Complexity, Topology and Dynamics in Quantum Matter ctd.qmat,
Julius-Maximilians-Universit\"at W\"urzburg, Am Hubland, D-97074 W\"urzburg, Germany}
\author{Pawe\l~Potasz}
\affiliation{Institute of Physics, Faculty of Physics, Astronomy and Informatics, Nicolaus Copernicus University, Grudziadzka 5, 87-100 Toru\'n, Poland}
\author{Ewelina M. Hankiewicz}
\affiliation{
Institute for Theoretical Physics and Astrophysics,
and W\"urzburg-Dresden Cluster of Excellence on Complexity, Topology and Dynamics in Quantum Matter ctd.qmat,
Julius-Maximilians-Universit\"at W\"urzburg, Am Hubland, D-97074 W\"urzburg, Germany}

\date{\today}

\begin{abstract}
The search for scalable, electrically controlled qubits remains a central challenge in quantum technology. We introduce gate-defined moiré quantum dots as a promising platform for valley-based qubits. Moiré engineering resolves the central conflict of valley physics: momentum-space separation protects the states, while the enlarged moiré length scale allows smooth gates to mix them controllably. Dot geometry and confinement strength program valley hybridization, while a displacement field controls detuning, providing two noncommuting electrostatic axes for qubit control.
\end{abstract}

\maketitle
\textit{Introduction.---}
Quantum computing has attracted enormous interest because it uses quantum
superposition and entanglement to process information in ways unavailable to
classical computers~\cite{feynman2018simulating,deutsch1985quantum,shor1994algorithms,grover1996fast, preskill2018quantum}.  Its basic unit is the quantum bit, or qubit: a coherent
quantum state that can be prepared, manipulated, and measured as a
superposition of the logical states \(\ket{0}\) and \(\ket{1}\).  Many
physical platforms are being developed for qubits, including superconducting
circuits, trapped ions, neutral atoms, defects in solids, and semiconductor
quantum dots~\cite{clarke2008superconducting,monroe2021programmable,
browaeys2020many,awschalom2018quantum,
Loss1998QD,petta2005coherent,kawakami2014electric,ciorga2001readout}.  In quantum dots, the
most established qubit uses the spin of a confined electron or hole.  Spin
qubits benefit from long coherence times, but achieving fast, fully electrical,
scalable, and reproducible control remains challenging.  This motivates the
search for other gate-controlled quantum degrees of freedom.

Valley-based qubits provide one such alternative
\cite{Culcer2012ValleyQC,Schoenfield2017valleyControl,
Penthorn2019FastValley,Kormanyos2014TMDQubits}.  In multivalley crystalline
materials such as silicon, graphene, and transition-metal dichalcogenides, the
electronic band structure contains multiple symmetry-related energy extrema at nonzero momenta,
called valleys, which can be used as an additional quantum degree of freedom.
Valley states are appealing because their separation in momentum space can
protect them from uncontrolled mixing, while their electronic wave functions
can still be influenced by electrostatic gates. 

Existing valley platforms,
however, reveal competition between protection and control.  In silicon quantum dots, both electrical tuning of the valley splitting and
coherent valley rotations have been demonstrated
~\cite{goswami2007controllable,Schoenfield2017valleyControl,Penthorn2019FastValley}, but the valley
splitting and mixing are tied to the microscopic structure of atomically sharp
interfaces~\cite{Culcer2010MultivalleySi,Culcer2012ValleyQC,
Tahan2014RelaxationSi,Zimmerman2017ValleyPhase,
Bourdet2018TunableSpinValley,
friesen2007valley,friesen2010theory,dodson2022valley,
paquelet2022atomic}. In bilayer graphene,
valley relaxation times exceeding $500$ ms have been measured, reflecting the
strong protection associated with large momentum separation, but controlled
valley mixing remains an open challenge for implementing coherent valley
rotations~\cite{Banszerus2021SpinValleyBLG,Garreis2024LongLived}.  Related
spin-valley qubit proposals in monolayer transition-metal dichalcogenide
(TMD) quantum dots exploit strong spin--orbit coupling and valley-dependent
orbital response, but intervalley coupling again relies on atomic-scale or
symmetry-breaking perturbations~\cite{Kormanyos2014TMDQubits,
Pawlowski2018ValleyMoS2,
Altintas2021SpinValley,Pawlowski2024Electrical,Sadecka2025MoSeWSeQD}.  These
examples show the central challenge: valleys are protected precisely because
they are hard to mix, while qubit control requires a controllable way to mix
them. 

Moir\'e superlattices have emerged as a powerful way to create highly tunable
solid-state systems~\cite{mak2022semiconductor}. In twisted two-dimensional materials, a small relative
rotation between layers produces a long-wavelength periodic potential whose
length scale is controlled by the twist angle.  This moir\'e potential can
strongly reshape the electronic bands, producing narrow minibands, correlated
states, and nontrivial band topology~\cite{naik2018ultraflatbands,
ruiz2019interlayer,Devakul2021Twisted,Regan2020,Tang2020,wang2020correlated,
cai2023signatures,zeng2023thermodynamic,park2023observation,xu2023observation}, and has recently been proposed as a platform for spin-based qubits~\cite{song2025twisted}.
Together with electrostatic gates and displacement fields, twist provides a
flexible route to engineering quantum states in van der Waals heterostructures.

In this manuscript, we propose gate-defined moir\'e quantum dots as a highly
promising platform for valley-based qubits.  The idea is simple but powerful:
replace atomic-scale valley coupling, which is difficult to control
electrostatically, with moir\'e-scale valley coupling that can be engineered
using smooth gates.  In a moir\'e miniband, the relevant valley extrema are
separated by a mini-Brillouin-zone momentum, much smaller than the momentum
separating atomic valleys.  Consequently, a smooth gate-defined quantum dot
can both confine carriers and provide sufficient momentum transfer to
hybridize the two moir\'e valleys.  This mechanism is not restricted to a
single material, but applies more broadly to the $K$-valley class of twisted
TMD homobilayers~\cite{Wu20219ContinuumModel, pan2020band,xu2026organizing}.  Using twisted WSe$_2$ as a concrete
example, we show that the lowest confined shell forms a moir\'e-valley
doublet whose hybridization is controlled by the dot radius and confinement
strength, while a displacement field independently controls the valley
detuning.  This realizes an electrically programmable moir\'e-valley
two-level system in which the protection-control balance is set by gate
design.

\begin{figure}[t]
    \centering
    \includegraphics[width=\linewidth]{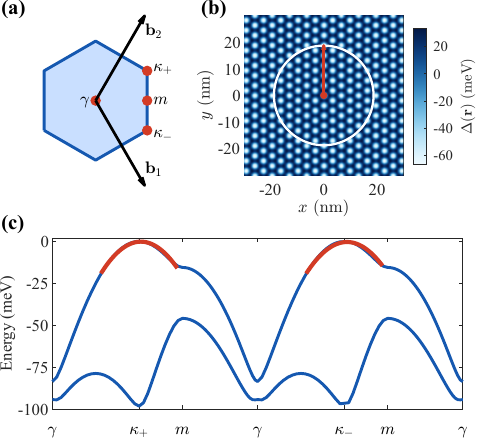}
    \caption{
Model ingredients for a gate-defined moir\'e quantum dot.
(a) Mini Brillouin zone showing the two moir\'e-valley maxima $\kappap$ and $\kappam$.
(b) Real-space moir\'e potential with a smooth Gaussian in-plane confinement of radius $R_{\rm QD}=W a_M$. The color scale shows the sum of the layer-diagonal moir\'e potentials, $\Delta(\mathbf r)=\Delta_1(\mathbf r)+\Delta_2(\mathbf r)$. (c) Top valence moir\'e bands at $\theta=5.08^\circ$. The red curves are parabolic fits near $\kappap$ and $\kappam$, whose curvature gives the hole effective mass $m^\ast_f=0.50m_e$.
}
    \label{fig:model}
\end{figure}

\textit{Continuum model and gate potentials.---}

We model a twisted WSe$_2$ homobilayer at $\theta=5.08^\circ$
($a_M\approx3.7$ nm), using the continuum Hamiltonian for the atomic
$K$ spin-valley sector.  The large valence-band spin--orbit splitting
locks the spin and atomic-valley characters near the band edge, allowing
the two time-reversed spin-valley sectors to be treated independently.
The time-reversed $K'$ sector provides the opposite-spin partner.
In the layer basis, the Hamiltonian has the form
\begin{equation}
H_K =
\begin{pmatrix}
-\dfrac{\hbar^2 |\mathbf k-\mathbf{\kappa}_+|^2}{2m^\ast}
+\Delta_1(\mathbf r)
&
\Delta_T(\mathbf r)
\\
\Delta_T^\dagger(\mathbf r)
&
-\dfrac{\hbar^2 |\mathbf k-\mathbf{\kappa}_-|^2}{2m^\ast}
+\Delta_2(\mathbf r)
\end{pmatrix}.
\label{eq:continuum}
\end{equation}
Figure~\ref{fig:model}(a)
shows the mini Brillouin zone and the two moir\'e-valley momenta
$\mathbf{\kappa}_\pm$. The density-functional-theory calibrated moir\'e potentials $\Delta_{1,2}$ and tunneling $\Delta_T$ are taken from Ref.~\cite{Devakul2021Twisted}. Fig.~\ref{fig:model}(b) shows the sum of the layer-diagonal moir\'e
potentials, \(\Delta(\mathbf r)=\Delta_1(\mathbf r)+\Delta_2(\mathbf r)\), and Fig.~\ref{fig:model}(c) shows the corresponding
top two valence moir\'e bands. Importantly, the
band maxima are located at $\mathbf{\kappa}_\pm$ and are approximately parabolic. 
\begin{figure}[t]
    \centering
    \includegraphics[width=\linewidth]{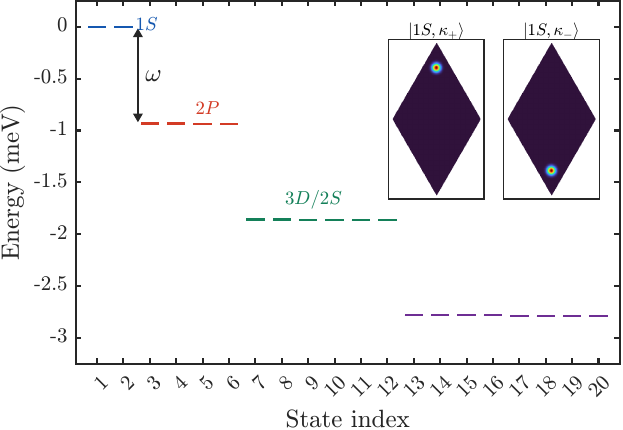}
    \caption{
    Confined orbital shell structure and moir\'e-valley character for a representative
    quantum dot with $W=50$, $V_0=100$ meV, and $D=0$.
    The single-particle spectrum forms oscillator-like orbitals, with a lowest
    $1S$ doublet separated from the $2P$ orbital by the spacing $\omega$.
    Insets show the momentum-space densities of the two $1S$ states, localized
    near the moir\'e valleys $\kappap$ and $\kappam$.
    }
    \label{fig:shell}
\end{figure}

\begin{figure*}[tbp]
    \centering
    \includegraphics[width=0.98\linewidth]{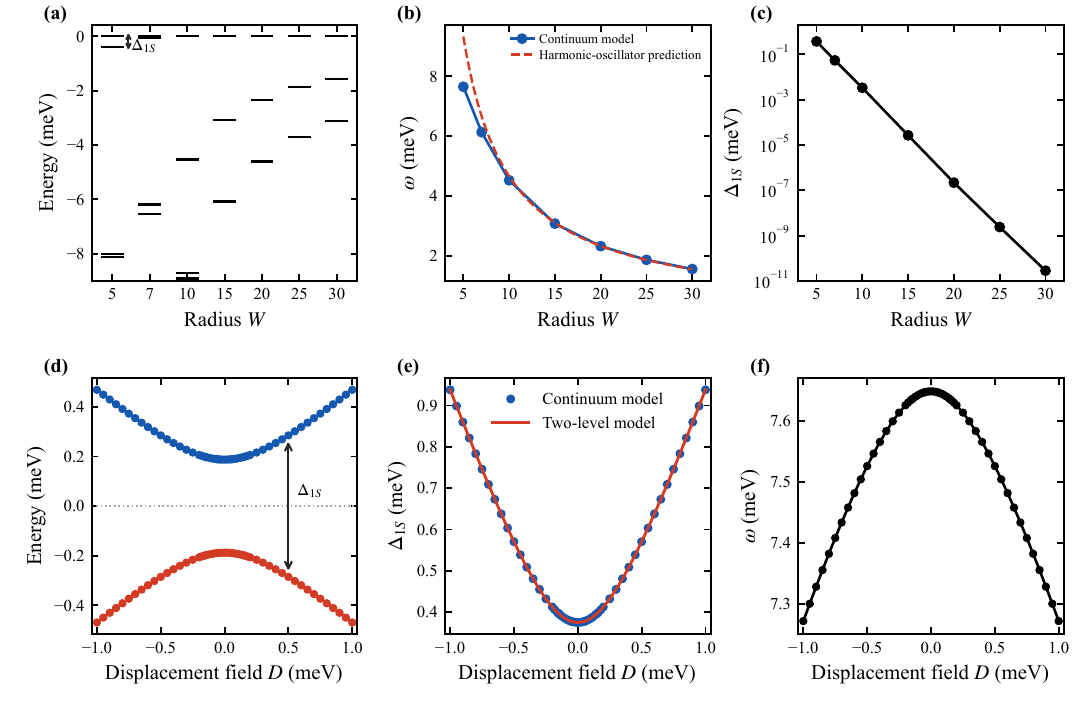}
    \caption{
     Gate control of the isolated $1S$ moir\'e-valley doublet.
    (a)--(c) Radius dependence at $\theta=5.08^\circ$, $V_0=100$ meV,
    and $D=0$: confined spectrum, single-particle orbital gap $\omega$,
    and zero-field $1S$ splitting $\Delta_{1S}=2|t|$. In (a), only the
    three lowest shells are shown. The dashed curve in (b) is the
    harmonic-oscillator prediction (see Supplemental Material), evaluated
    using the hole effective mass $m^\ast_f=0.50m_e$ extracted from the
    parabolic fits in Fig.~\ref{fig:model}(c).
    (d)--(f) Displacement-field dependence at the same twist angle for
    $W=5$ and $V_0=100$ meV: centered $1S$ levels, two-level splitting,
    and orbital gap $\omega$. The radius tunes the intervalley
    hybridization, while the displacement field tunes the valley detuning
    without closing the orbital gap.}
    \label{fig:control}
\end{figure*}
A gate-defined quantum dot is introduced through
\begin{equation}
H =
H_K+
V_{\rm QD}(\mathbf r)+\frac{D}{2}\sigma_z,
\label{eq:gated_hamiltonian}
\end{equation}
with
$V_{\rm QD}(\mathbf r)
=
V_0 e^{-r^2/R_{\rm QD}^2}\sigma_0$,
$R_{\rm QD}=W a_M$. The dot radius is indicated in Fig.~\ref{fig:model}(b).  Here
$\sigma_i$ is the $i$-th Pauli matrix which act in layer space, $D$ is the layer-asymmetric displacement field, $V_0$ is the confinement depth, and $W$ measures the dot radius in moir\'e lattice constants.  The confined states are expanded in the bulk moiré-band basis, following the projection strategy used in finite moir\'e geometries~\cite{saleem2026topological}. Numerical and implementation details can be found in the Supplemental Material.

\textit{Confined orbitals and the $1S$ doublet.---}
The physical role of the gate in Fig.~\ref{fig:model}(b) is transparent near the band maxima in Fig.~\ref{fig:model}(c).  The maxima are approximately parabolic, while the Gaussian confinement is locally harmonic; the confined hole spectrum therefore organizes into oscillator-like orbitals (see Supplemental Material).  Figure \ref{fig:shell} shows this structure for a representative large dot with $W=50$ ($R_{\rm QD}\simeq185$ nm).  The spectrum forms a lowest $1S$ doublet,
followed by a $2P$ orbital and higher excited orbitals.  The momentum-space densities in the insets show that the two $1S$ states originate from the two moiré-valley maxima $\kappap$ and $\kappam$.  We therefore use $\{\ket{1S,\kappap},\ket{1S,\kappam}\}$ as the effective two-level subspace.  The spacing $\omega$ from the $1S$ doublet to the lowest $2P$ state defines the orbital leakage scale. 

Figure~\ref{fig:control}(a)--(c) shows how the confined spectrum evolves with dot radius.  The spacing between the lowest $1S$ orbital and the excited $2P$ orbital decreases smoothly with increasing $W$, as expected for a larger quantum dot, but remains on the meV scale throughout the range shown. In contrast, the splitting within the $1S$ doublet is much more sensitive to dot size: it is hundreds of $\mu$eV for $W=5$, tens of $\mu$eV for $W=7$, and only a few $\mu$eV for $W=10$.  This behavior follows from the momentum-space structure of the confinement potential (See Supplemental Material). A smaller dot has a more localized real-space envelope and therefore a
broader momentum-space distribution. It consequently couples the two
moir\'e valleys $\kappap$ and $\kappam$ more strongly, whereas a larger
dot approaches the moir\'e-momentum-conserving limit and recovers nearly
valley-resolved states.

\textit{Effective valley Hamiltonian.---}
The lowest orbital doublet defines a two-dimensional moir\'e-valley subspace,
spanned by the $1S$ states associated with $\kappap$ and $\kappam$.  After
projecting Eq.~\eqref{eq:gated_hamiltonian} onto this subspace and dropping an
overall energy shift, the effective Hamiltonian can be written as
\begin{equation}
H_{1S}= \boldsymbol{\tau}\cdot \mathbf d,
\label{eq:heff_general}
\end{equation}
where the Pauli matrices $\boldsymbol{\tau}$ acts in the $\{\kappap,\kappam\}$ valley basis.  The
vector $\mathbf d$ is set by two projected matrix elements: the off-diagonal
hybridization
\begin{equation}
t=\mel{1S,\kappap}{V_{\rm QD}}{1S,\kappam},
\label{eq:t_def}
\end{equation}
generated by the finite in-plane confinement, and the valley detuning
\begin{equation}
\epsv =
\frac{D}{2}\left(\eta_+-\eta_-\right),
\qquad
\eta_\pm=\mel{1S,\kappa_\pm}{\sigma_z}{1S,\kappa_\pm},
\label{eq:epsv_def}
\end{equation}
generated by the displacement field.  Thus
\begin{equation}
\mathbf d =
\left(
|t|,
0,
\frac{\epsv}{2}
\right).
\label{eq:d_vector}
\end{equation}
In this valley basis, $|t|$ sets the hybridization axis and is controlled by
the dot radius and potential depth, while $d_z=\epsv/2$ sets the valley-detuning axis and
is controlled by the displacement field. Microscopically, $t$ requires momentum
transfer between $\kappap$ and $\kappam$ and is therefore governed by the
Fourier width of the Gaussian confinement potential, as derived in the
Supplemental Material.

Diagonalizing Eq.~\eqref{eq:heff_general} gives the splitting between the two
$1S$ states,
\begin{equation}
\Delta_{1S}
=\sqrt{4|t|^2+\epsv^2}.
\label{eq:splitting}
\end{equation}
The corresponding eigenstates, which we denote by $\ket{\psi_1}$ and $\ket{\psi_2}$ in order of increasing energy within the
$1S$ doublet, are controlled by the direction of $\mathbf d$ on the
pseudospin Bloch sphere.  In the detuned limit $|\epsv|\gg |t|$, they approach the valley-resolved states $\ket{1S,\kappap}$ and $\ket{1S,\kappam}$.  At zero displacement field, $\epsv=0$, they become bonding and antibonding superpositions of the two moir\'e-valley states.

Figures~\ref{fig:control}(d) and \ref{fig:control}(e)  show the displacement-field response of the
lowest doublet.  After subtracting the mean energy of the two $1S$ levels, the
spectrum evolves symmetrically with $D$, as expected for a gate-controlled
valley detuning.  The resulting splitting follows Eq.~\eqref{eq:splitting},
with the zero-field gap (equal to $2|t|$)  set by confinement-induced hybridization and the
field-dependent part set by $\epsv(D)$.  The orbital spacing $\omega$ shown in Fig.~\ref{fig:control}(f) remains
large throughout the sweep, confirming that the displacement field tunes the
$1S$ valley Hamiltonian without strongly mixing with excited orbitals. 

\begin{figure}[tbp]
    \centering
    \includegraphics[width=0.98\linewidth]{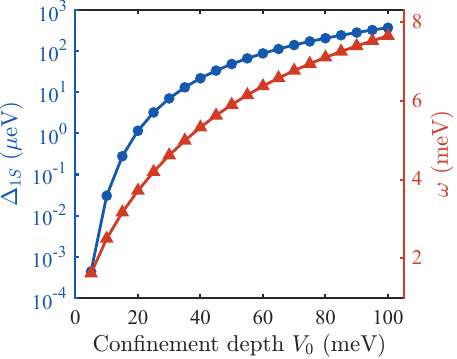}
    \caption{
    Confinement-depth control at fixed $W=5$ and $D=0$. The blue data shows the $1S$ splitting $\Delta_{1S}=2|t|$ versus  confinement depth plotted on a log scale. The red data shows the orbital spacing $\omega$ versus confinement depth.}
    \label{fig:depth}
\end{figure}

The remaining control knob is the confinement potential depth.  Figure~\ref{fig:depth}
shows that increasing $V_0$ enhances the zero-displacement-field splitting of the $1S$
doublet.  In the effective Hamiltonian this corresponds to an increase of the
intervalley matrix element $|t|$.  The dependence on $V_0$ has two related
contributions.  First, the confinement potential itself enters the matrix
element, so increasing its amplitude directly increases the scale of $t$.
Second, a deeper dot produces a more localized $1S$ envelope in real space,
and hence a broader envelope in momentum space.  This increases the weight of
the confined state at the moir\'e-valley momentum transfer needed to connect
$\kappap$ and $\kappam$.  At the same time, the orbital spacing $\omega$
also grows, showing that stronger intervalley coupling can be obtained while
maintaining separation from the higher-energy orbitals.

Taken together, Figs.~\ref{fig:control} and \ref{fig:depth} establish the
control structure summarized by Eq.~\eqref{eq:d_vector}: the dot radius and
confinement depth tune the intervalley hybridization $|t|$, while the
displacement field tunes the valley detuning $\epsv$.  These two
noncommuting electrostatic controls define the accessible axes of the
moiré-valley pseudospin. 

\textit{Discussion and outlook.---}
The results above show that a gate-defined moir\'e quantum dot can realize a
two-level valley system whose Hamiltonian is programmed electrostatically.
The confinement radius and depth tune the intervalley matrix element $|t|$,
while the displacement field independently tunes the detuning $\epsv$,
making the balance between valley protection and control programmable.
Beyond these electrostatic controls, the twist angle provides an additional
design parameter by setting the moir\'e length scale and, consequently, the
momentum-space separation between $\kappap$ and $\kappam$.  Although our
calculations focus on WSe$_2$ at a fixed twist angle, the underlying mechanism
applies more broadly to $K$-valley twisted TMD homobilayers with moir\'e-band
extrema separated by mini-Brillouin-zone momenta.  Our construction therefore
belongs to the broader classification of moir\'e materials based on the
momentum and orbital character of the parent-material band
edges~\cite{xu2026organizing}.

Within a fixed spin-valley sector, this independent control supports two natural operating modes. In a detuned regime, $|\epsv|\gg |t|$, the eigenstates are approximately the valley states $\ket{1S,\kappap}$ and $\ket{1S,\kappam}$; an ac modulation of the confinement potential depth then drives the hybridization term $t(t)$ and can rotate the moir\'e-valley pseudospin.  Conversely, near $D=0$ ($|\epsv|=0$), the eigenstates $\ket{\psi_1}$ and $\ket{\psi_2}$ are bonding and antibonding superpositions of the two valley states, and an ac displacement field provides the noncommuting drive. Which mode is optimal is therefore a device-design question: one may choose either valley-resolved states or hybridized eigenstates as the working two-level system.

Because the control mechanism is electrical, charge noise will be an important
consideration in choosing the operating regime.  Near the center of the
avoided crossing, the qubit splitting is first-order insensitive to slow
displacement-field fluctuations, although it remains sensitive to fluctuations
of the confinement potential.  In the strongly detuned regime, the situation
is reversed: sensitivity to confinement-induced hybridization noise is
suppressed, while displacement-field noise becomes more important.  The
device design and operating point can therefore be chosen according to the
dominant experimental noise source.  Quantitative coherence and relaxation
times require device-specific gate-noise and phonon spectra and remain for
future investigation~\cite{Reed2016Reduced,Wang024ValleyRelax}.

The two-state system discussed here corresponds to one atomic spin-valley
sector.  Time-reversal symmetry generates a degenerate partner doublet in the
opposite spin-valley sector, which is not included in
Eq.~\eqref{eq:heff_general}.  This additional copy does not alter the control
mechanism within either sector, but a single-qubit implementation would
require selecting one of them, for example using a weak magnetic field or
spin-valley-selective initialization.  Such selection may also arise from
interactions: in some twisted TMD homobilayers at relatively small twist
angles and near half filling of the top valence moir\'e band,
electron--electron interactions can drive spontaneous spin-valley
polarization~\cite{Devakul2021Twisted,abouelkomsan2024band,
zhang2021electronic}, thereby lifting the degeneracy between the two
time-reversed spin-valley sectors.

The relevant length and energy scales are close to those of existing
gate-defined TMD quantum dots.  For $\theta=5.08^\circ$, $W=5$ corresponds to $R_{\rm QD}\simeq 19$ nm, or a Gaussian diameter of order $40$ nm.  This length scale is experimentally realistic: gate-defined WSe$_2$ quantum dots with effective diameters below $60$ nm  have been reported~\cite{davari2020gate,boddison2021gate}. The calculated \(1S\) splittings reach several hundred
\(\mu{\rm eV}\), while the spacing to the \(2P\) shell remains on the meV
scale. Thus the two-level splittings correspond to electrically addressable
frequencies from tens to hundreds of gigahertz, while remaining well separated
from the higher-energy orbitals. This establishes laterally gated moir\'e quantum dots as a platform for electrostatically engineered valley-based two-level systems, where hybridization, detuning, and the operating basis are controlled by gate design.

\begin{acknowledgments}
\textit{Acknowledgments.---}
Y.S. would like to thank Bj\"orn Trauzettel,
Pawe{\l} Hawrylak, Daniel Miravet, and Alina Wania Rodrigues for useful discussions. We acknowledge financial support by the Deutsche Forschungsgemeinschaft (DFG, German Research Foundation) through the Würzburg-Dresden Cluster of Excellence ctd.qmat – Complexity, Topology and Dynamics in Quantum Matter (EXC 2147, project-id 390858490).
\end{acknowledgments}

\bibliographystyle{apsrev4-2}
\bibliography{Main}

\end{document}


\title{Supplemental Material}

\author{Yasser Saleem}
\affiliation{
Institute for Theoretical Physics and Astrophysics,
and W\"urzburg-Dresden Cluster of Excellence on Complexity, Topology and Dynamics in Quantum Matter ctd.qmat,
Julius-Maximilians-Universit\"at W\"urzburg, Am Hubland, D-97074 W\"urzburg, Germany}
\author{Pawe\l~Potasz}
\affiliation{Institute of Physics, Faculty of Physics, Astronomy and Informatics, Nicolaus Copernicus University, Grudziadzka 5, 87-100 Toru\'n, Poland}
\author{Ewelina M. Hankiewicz}
\affiliation{
Institute for Theoretical Physics and Astrophysics,
and W\"urzburg-Dresden Cluster of Excellence on Complexity, Topology and Dynamics in Quantum Matter ctd.qmat,
Julius-Maximilians-Universit\"at W\"urzburg, Am Hubland, D-97074 W\"urzburg, Germany}

\date{\today}
\maketitle

\section{Moir\'e-band basis and confinement matrix elements}

We use the continuum Hamiltonian defined in the main text, with parameters from Ref.~\cite{Devakul2021Twisted}, and diagonalize the unconfined moir\'e problem at crystal momentum $\mathbf k$ in the mini Brillouin zone.  The resulting Bloch states are written as
\begin{equation}
\Phi_{p\mathbf k}(\mathbf r)=
\frac{1}{\sqrt A}
\sum_{\mathbf G}
 e^{i(\mathbf k+\mathbf G)\cdot \mathbf r}
\begin{pmatrix}
A^1_{p,\mathbf k+\mathbf G}\\
A^2_{p,\mathbf k+\mathbf G}
\end{pmatrix},
\label{eq:SM_bulk_wf}
\end{equation}
where $p$ labels the moir\'e band, $\mathbf G$ are moir\'e reciprocal lattice vectors, and $A$ is the system area.  Confined states are expanded as in the finite-geometry projection strategy of Ref.~\cite{saleem2026topological},
\begin{equation}
\ket{\psi_s}=\sum_{p,\mathbf k} C^s_{p\mathbf k}\ket{\Phi_{p\mathbf k}} .
\label{eq:SM_dot_expansion}
\end{equation}
The lateral dot potential is taken to be
\begin{equation}
V_{\rm QD}(\mathbf r)=V_0 e^{-r^2/R_{\rm QD}^2}\sigma_0,
\qquad R_{\rm QD}=W a_M .
\label{eq:SM_gaussian}
\end{equation}
Its matrix elements in the moir\'e-band basis are
\begin{align}
&\mel{\Phi_{p\mathbf k}}{V_{\rm QD}}{\Phi_{p'\mathbf k'}} 
\nonumber\\
&=\frac{1}{A}
\sum_{\alpha=1,2}\sum_{\mathbf G,\mathbf G'}
\left(A^{\alpha}_{p,\mathbf k+\mathbf G}\right)^*
A^{\alpha}_{p',\mathbf k'+\mathbf G'}
\widetilde V_{\rm QD}(\mathbf k'-\mathbf k+\mathbf G'-\mathbf G),
\label{eq:SM_V_matrix}
\end{align}
with the two-dimensional Fourier transform
\begin{align}
\widetilde V_{\rm QD}(\mathbf Q)
&=\int d^2r\,e^{-i\mathbf Q\cdot\mathbf r}V_0e^{-r^2/R_{\rm QD}^2}\nonumber\\
&=\pi R_{\rm QD}^2V_0e^{-Q^2R_{\rm QD}^2/4}.
\label{eq:SM_gaussian_FT}
\end{align}
Equation~\eqref{eq:SM_V_matrix} shows explicitly why smooth confinement has a different effect in moir\'e and atomic valley systems.  For atomic valleys, the required momentum transfer is of order $1/a_0$ and the Gaussian Fourier component is exponentially small unless the perturbation is atomically localized.  For moir\'e valleys, the transfer momentum is of order $1/a_M$, so a dot containing only a finite number of moir\'e unit cells can generate an appreciable intervalley matrix element.

\section{Effective two-level Hamiltonian}

We now project the gated Hamiltonian onto the lowest confined shell.  In the
absence of displacement field, the relevant low-energy subspace is spanned by
the two $1S$ moir\'e-valley states
\begin{equation}
    \left\{
    \ket{1S,\kappap},
    \ket{1S,\kappam}
    \right\}.
\end{equation}
In this basis, and after subtracting the average energy of the doublet, the
most general projected Hamiltonian can be written as
\begin{equation}
    H_{1S}
    =
    \begin{pmatrix}
        \epsv/2 & t \\
        t^\ast & -\epsv/2
    \end{pmatrix},
\label{eq:SM_H1S_matrix}
\end{equation}
where
\begin{equation}
    t
    =
    \mel{1S,\kappap}{V_{\rm QD}}{1S,\kappam}
\label{eq:SM_t_def}
\end{equation}
is the confinement-induced intervalley matrix element.  The complex phase of
$t$  can be absorbed into
the relative phase of the two valley basis states.  We therefore choose the
valley basis such that 
\begin{equation}
    H_{1S}
    =
    |t|\tau_x+\frac{\epsv}{2}\tau_z,
\label{eq:SM_H1S}
\end{equation}
or equivalently
\begin{equation}
    H_{1S}=\boldsymbol{\tau}\cdot\mathbf d,
    \qquad
    \mathbf d=
    \left(
        |t|,
        0,
        \epsv/2
    \right).
\label{eq:SM_dvector}
\end{equation}

The displacement field enters the microscopic Hamiltonian as a layer-asymmetric
perturbation,
\begin{equation}
    H_D=\frac{D}{2}\sigma_z ,
\label{eq:SM_HD}
\end{equation}
where $\sigma_z$ acts in layer space.  Projecting this term into the $1S$
moire-valley doublet gives the valley detuning
\begin{equation}
    \epsv(D)
    =
    \frac{D}{2}
    \left(
        \eta_+-\eta_-
    \right),
    \qquad
    \eta_\pm
    =
    \mel{1S,\kappa_\pm}{\sigma_z}{1S,\kappa_\pm}.
\label{eq:SM_detuning}
\end{equation}

\section{Parabolic-band limit and oscillator shells}

The shell structure of the laterally confined states can be understood from
the projected single-band problem near a moir\'e-valley maximum.  We write the
Hamiltonian as
\begin{equation}
    H = H_K + V_{\rm QD}(\mathbf r),
\end{equation}
where \(H_K\) is the unconfined moir\'e continuum Hamiltonian and
\(V_{\rm QD}\) is the smooth lateral gate potential.  The eigenstates of
\(H_K\) are moir\'e Bloch states,
\begin{equation}
    H_K\ket{\Phi_{p,\mathbf k}}
    =
    \epsilon_{p,\mathbf k}\ket{\Phi_{p,\mathbf k}},
\end{equation}
with real-space wave functions
\begin{equation}
    \Phi_{p,\mathbf k}(\mathbf r)
    =
    e^{i\mathbf k\cdot\mathbf r}
    u_{p,\mathbf k}(\mathbf r).
\end{equation}
Here \(p\) labels the moir\'e band and \(\epsilon_{p,\mathbf k}\) is the
corresponding unconfined moir\'e-band energy.  For the low-energy confined
states considered here, it is sufficient to keep a single isolated valence
band, and we suppress the band index below:
\begin{equation}
    \epsilon_{\mathbf k}\equiv \epsilon_{p,\mathbf k},
    \qquad
    \ket{\Phi_{\mathbf k}}\equiv\ket{\Phi_{p,\mathbf k}} .
\end{equation}

A confined eigenstate is expanded in this unconfined moir\'e-band basis as
\begin{equation}
    \ket{\varphi_s}
    =
    \sum_{\mathbf k} C^s_{\mathbf k}\ket{\Phi_{\mathbf k}} .
\end{equation}
Projecting \(H\ket{\varphi_s}=E_s\ket{\varphi_s}\) onto
\(\bra{\Phi_{\mathbf k}}\) gives
\begin{equation}
    \epsilon_{\mathbf k} C^s_{\mathbf k}
    +
    \sum_{\mathbf k'}
    C^s_{\mathbf k'}
    \bra{\Phi_{\mathbf k}}
    V_{\rm QD}(\mathbf r)
    \ket{\Phi_{\mathbf k'}}
    =
    E_s C^s_{\mathbf k}.
\label{eq:projected_schrodinger}
\end{equation}
The confinement matrix element contains the Fourier transform of the gate
potential and the moir\'e Bloch-state form factor,
\begin{equation}
    \bra{\Phi_{\mathbf k}}
    V_{\rm QD}
    \ket{\Phi_{\mathbf k'}}
    =
    \frac{1}{A}
    \sum_{\Delta\mathbf G}
    F_{\Delta\mathbf G}(\mathbf k,\Delta\mathbf k)
    \widetilde V_{\rm QD}(\Delta\mathbf k+\Delta\mathbf G),
\label{eq:confinement_matrix_element}
\end{equation}
where
\begin{equation}
    \Delta\mathbf k=\mathbf k'-\mathbf k
\end{equation}
and
\begin{equation}
    F_{\Delta\mathbf G}(\mathbf k,\Delta\mathbf k)
    =
    \bra{u_{\mathbf k}}
    e^{-i\Delta\mathbf G\cdot\mathbf r}
    \ket{u_{\mathbf k+\Delta\mathbf k}} .
\end{equation}

The important small parameter is not \(\Delta\mathbf k\) or
\(\Delta\mathbf G\) separately, but the combined momentum transfer
\begin{equation}
    \boldsymbol\lambda
    =
    \Delta\mathbf k+\Delta\mathbf G .
\end{equation}
For a Gaussian confinement,
\begin{equation}
    \widetilde V_{\rm QD}(\mathbf q)
    \propto
    e^{-q^2R_{\rm QD}^2/4},
\end{equation}
so the dominant terms in Eq.~\eqref{eq:confinement_matrix_element} are those
that minimize $|\boldsymbol\lambda|$.  In the moir\'e Brillouin zone this
requires keeping the $\Delta\mathbf G=0$ channel together with the first
shell of moir\'e reciprocal lattice vectors.  The second shell always gives a
larger value of $|\Delta\mathbf k+\Delta\mathbf G|$ since $\textbf{k},\textbf{k}'$ are confined to the first Brillouin zone. We therefore define the minimum-transfer set
\begin{equation}
    \mathcal S_{\rm min}
    =
    \left\{
    \Delta\mathbf G:
    |\Delta\mathbf k+\Delta\mathbf G|
    \text{ is minimal}
    \right\},
\end{equation}
which, in practice, is contained in the zeroth and first reciprocal-lattice
shells.  For generic momenta there is a unique vector
\(\Delta\mathbf G_{\rm min}\) that minimizes
\(|\Delta\mathbf k+\Delta\mathbf G|\).  Non-unique choices occur only on
Brillouin-zone boundaries, where two reciprocal-lattice images are equally
close.  We ignore these boundary cases in the approximate mapping below.  Expanding the form factors
locally around the valley maximum and retaining only the leading term gives
\begin{equation}
    F_{\Delta\mathbf G}(\mathbf k,\Delta\mathbf k)
    =
    1
    +
    O(\lambda).
    \label{eq:0thOrderFormFactor}
\end{equation}
The leading confinement matrix element is then
\begin{equation}
    \bra{\Phi_{\mathbf k}}
    V_{\rm QD}
    \ket{\Phi_{\mathbf k'}}
    \approx
    \frac{1}{A}
    \widetilde V_{\rm QD}
    \left(
        \Delta\mathbf k+\Delta\mathbf G_{\rm min}
    \right).
\label{eq:leading_moire_matrix_element}
\end{equation}
We note, higher-order terms in \(\boldsymbol\lambda\) contain the Berry connection and
quantum geometric tensor of the moir\'e band, and therefore generate
geometric corrections to the simple effective-mass description.  In the
parameter regime studied here, the numerically obtained low-energy spectra
are well described by the leading-order shell structure, and we do not
resolve these geometric corrections.  Understanding when such terms become
important is an interesting direction beyond the scope of the present work.

Close to a given moir\'e-valley maximum, we write
\begin{equation}
    \mathbf k=\boldsymbol{\kappa}_{\tau}+\mathbf q,
    \qquad
    \tau=\pm ,
\end{equation}
and similarly
\(\mathbf k'=\boldsymbol{\kappa}_{\tau}+\mathbf q'\).  The reciprocal vector
\(\Delta\mathbf G_{\rm min}\) selects the nearest reciprocal-lattice image of
\(\mathbf q'\).  Equivalently, one may introduce unfolded local momenta for
which
\begin{equation}
    \Delta\mathbf q+\Delta\mathbf G_{\rm min}
    =
    \overline{\mathbf q}'-\overline{\mathbf q}.
\end{equation}
We use these unfolded coordinates below and drop the overbar for simplicity.
The top valence band near either maximum is approximately parabolic,
\begin{equation}
    \epsilon_{\tau}(\mathbf q)
    \approx
    \epsilon_{\tau}^{0}
    -
    \frac{\hbar^2 q^2}{2m^\ast_f},
    \label{eq:ParabolicMaxima}
\end{equation}
where \(m^\ast_f\) is the effective mass extracted from the moir\'e valence-band
maximum at \(\boldsymbol{\kappa}_{\tau}\).

We now take the continuum limit of the momentum grid,
\begin{equation}
    \sum_{\mathbf q'}
    \longrightarrow
    \frac{A}{(2\pi)^2}
    \int d^2q',
\end{equation}
where \(A\) is the real-space normalization area.   Equivalently,
\begin{equation}
    \frac{1}{A}\sum_{\mathbf q'}
    \longrightarrow
    \int\frac{d^2q'}{(2\pi)^2}.
\end{equation}
The projected equation therefore becomes
\begin{equation}
\begin{aligned}
    \left(
        \epsilon_{\tau}^{0}
        -
        \frac{\hbar^2q^2}{2m^\ast_f}
    \right)
    C^s(\mathbf q)
    &{}+
    \int\frac{d^2q'}{(2\pi)^2}
    \\
    &\quad\times
    \widetilde V_{\rm QD}(\mathbf q-\mathbf q')
    C^s(\mathbf q')
    =
    E_s C^s(\mathbf q).
\end{aligned}
\label{eq:continuum_projected_equation}
\end{equation}

To transform this equation to real space, we define
\begin{equation}
    \chi_s(\mathbf r)
    =
    \int\frac{d^2q}{(2\pi)^2}
    C^s(\mathbf q)e^{i\mathbf q\cdot\mathbf r}.
\label{eq:SM_envelope}
\end{equation}
The confinement term in
Eq.~\eqref{eq:continuum_projected_equation} is a momentum-space convolution.
Using the convolution theorem,
\begin{equation}
    \int\frac{d^2q'}{(2\pi)^2}
    \widetilde V_{\rm QD}(\mathbf q-\mathbf q')
    C^s(\mathbf q')
    \quad\longleftrightarrow\quad
    V_{\rm QD}(\mathbf r)\chi_s(\mathbf r).
\label{eq:SM_convolution}
\end{equation}
Likewise, multiplication by \(q^2\) in momentum space becomes
\(-\nabla^2\) in real space.  Equation~\eqref{eq:continuum_projected_equation}
therefore becomes
\begin{equation}
    \left[
        \epsilon_{\tau}^{0}
        +
        \frac{\hbar^2\nabla^2}{2m^\ast_f}
        +
        V_{\rm QD}(\mathbf r)
    \right]
    \chi_s(\mathbf r)
    =
    E_s\chi_s(\mathbf r).
\label{eq:SM_real_space_equation}
\end{equation}

For the Gaussian gate,
\begin{equation}
    V_{\rm QD}(\mathbf r)
    =
    V_0e^{-r^2/R_{\rm QD}^2}
    \approx
    V_0-\frac{V_0}{R_{\rm QD}^2}r^2
\end{equation}
near the center of the dot.  Since we consider holes near a valence-band
maximum, it is convenient to measure the confinement energy downward from
the local maximum,
\begin{equation}
    \mathcal E_s
    =
    \epsilon_{\tau}^{0}+V_0-E_s.
\end{equation}
After removing the constant energy shift,
Eq.~\eqref{eq:SM_real_space_equation} becomes
\begin{equation}
    \left[
        -\frac{\hbar^2\nabla^2}{2m^\ast_f}
        +
        \frac{V_0}{R_{\rm QD}^2}r^2
    \right]
    \chi_s(\mathbf r)
    =
    \mathcal E_s\chi_s(\mathbf r).
\label{eq:SM_hole_oscillator}
\end{equation}
This is the two-dimensional harmonic oscillator,
with
\begin{equation}
    \omega
    =
    \sqrt{
        \frac{2V_0}{m^\ast_f R_{\rm QD}^2}
    }.
\end{equation}
The corresponding shell energies are
\begin{equation}
    E_{n_x,n_y}
    =
    \hbar\omega(n_x+n_y+1),
\end{equation}
up to an overall energy shift and corrections from band nonparabolicity, the
finite range of the Gaussian potential, and the neglected higher-order
geometric terms.

This leading-order picture explains the numerical shell sequence observed in
the continuum calculation: a lowest \(1S\) shell, followed by a \(2P\) shell,
then higher \(2S/3D\)-like states.  Each orbital shell carries the two
moir\'e-valley labels \(\kappap\) and \(\kappam\).  The lowest \(1S\) doublet
therefore provides the effective valley subspace used in the main text, while
the spacing to the \(2P\) shell defines the orbital leakage scale \(\omega\).

\section{Numerical convergence}
The confined-state calculation is controlled by three numerical parameters:
the momentum-space grid \(N_1=N_2\), the reciprocal-lattice cutoff \(N_G\),
and the number \(N_B\) of retained moir\'e bands.  We test convergence with
respect to all three parameters for the smallest dot considered in the main
text, \(W=5\), at \(\theta=5.08^\circ\), \(V_0=100\) meV, and \(D=0\).
This provides the most stringent test of the band truncation because the
strong confinement produces the largest intervalley hybridization.

\begin{figure*}[tbp]
    \centering
    \includegraphics[width=0.8\linewidth]{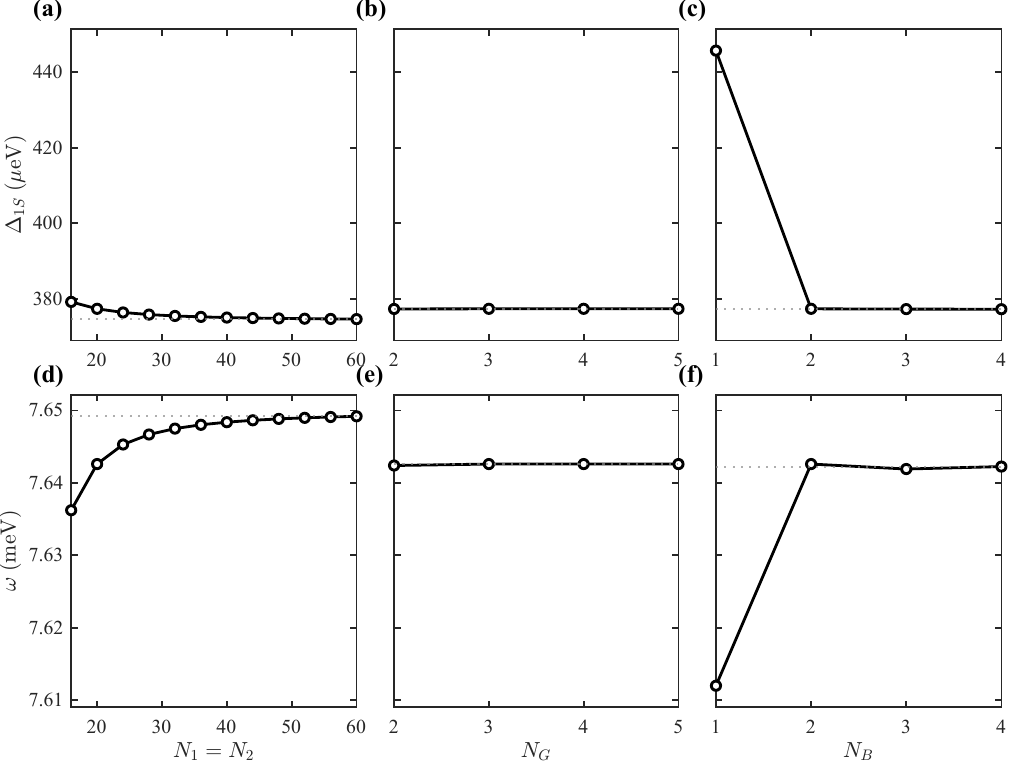}
    \caption{
    Numerical convergence of the zero-field \(1S\) splitting
    \(\Delta_{1S}\) (top row) and orbital spacing \(\omega\) (bottom row)
    for \(W=5\), \(V_0=100\) meV, and \(\theta=5.08^\circ\).
    Panels (a) and (d) vary the momentum-space grid \(N_1=N_2\);
    panels (b) and (e) vary the reciprocal-lattice cutoff \(N_G\);
    and panels (c) and (f) vary the number \(N_B\) of retained moir\'e
    bands.  The remaining parameters are held fixed in each sweep.
    Dotted horizontal lines indicate the converged values.
    }
    \label{fig:SM_convergence}
\end{figure*}

Figure~\ref{fig:SM_convergence} demonstrates convergence of both
\(\Delta_{1S}\) and \(\omega\).  For the \(W=5\) and \(W=7\) radius points
in the main text, we use \(N_1=N_2=60\), \(N_G=4\), and \(N_B=2\).
For the larger dots, we use \(N_G=4\), \(N_B=1\), and momentum-space grids
of \(N_1=N_2=140\), which are sufficient to converge both
quantities.

\bibliographystyle{apsrev4-2}
\bibliography{Main}